# Understanding High Dimensional Spaces through Visual Means Employing Multidimensional Projections

**Haseeb Younis**

University College Cork, Ireland, 121126205@umail.ucc.ie, https://orcid.org/**0000-0003-4093-3027**

**Paul Trust**

University College Cork, Ireland, 120222601@umail.ucc.ie, https://orcid.org/0000-0003-0177-1096

**Rosane Minghim**

University College Cork, Ireland, r.minghim@cs.ucc.ie, https://orcid.org/0000-0002-4799-8774

**Abstract:** Data visualisation helps understanding data represented by multiple variables, also called features, stored in a large matrix where individuals are stored in lines and variable values in columns. These data structures are frequently called multidimensional spaces. A large set of mathematical tools, named frequently as multidimensional projections, aim to map such large spaces into 'visual spaces', that is, to 2 or 3 dimensions, where the aspect of that space can be visualised. While the final product is intuitive in that proximity between points - or iconic representation of points - indicate similarity relationships in the original space, understanding the formulation of the projection methods many times escapes researchers. In this paper, we illustrate ways of employing the visual results of multidimensional projection algorithms to understand and fine-tune the parameters of their mathematical framework. Some of the common mathematical common to these approaches are Laplacian matrices, Euclidian distance, Cosine distance, and statistical methods such as Kullback-Leibler divergence, employed to fit probability distributions and reduce dimensions. Two of the relevant algorithms in the data visualisation field are t-distributed stochastic neighbourhood embedding (t-SNE) and Least-Square Projection (LSP). These algorithms can be used to understand several ranges of mathematical functions including their impact on datasets. In this article, mathematical parameters of underlying techniques such as Principal Component Analysis (PCA) behind t-SNE and mesh reconstruction methods behind LSP are adjusted to reflect the properties afforded by the mathematical formulation. The results, supported by illustrative methods of the processes of LSP and t-SNE, are meant to inspire students in understanding the mathematics behind such methods, in order to apply them in effective data analysis tasks in multiple applications.

## Introduction

Data Visualization is concerned with displaying large amounts of data to aid discovery of important patterns and gain knowledge. It helps to understand the data distribution and refine it by using complex mathematical functions. Based on mathematical formulations, some visualization tools can interpret provide clear representations of data aspects even in complex set ups, such as high amounts of describing variables. It is important to visualize the high dimensional matrices to understand the behaviour unstructures data. In mathematics, visualization has a lengthy history, with a long list of renowned mathematicians adopting or expressly advocating it. It can be a useful tool for delving into mathematical difficulties and explaining mathematical concepts and relationships. When working with large data sets, visualization can help to simplify things [1]. Conventional visualization methods such as scatter plots, histograms, and parallel coordinate plots are still widely used in current visual analytics [2] tasks. The disadvantage of these visualization approaches is that they can only visualize small number of data variables at a time, preventing them from being used to big, high-dimensional data sets.

To understand the impact of mathematical operations in using visualization, a reduced number of dimensions can perform best, as large numbers of dimensions are difficult to interpret. Exploring a low-dimensional embedding of the multi-dimensional data is a typical approach to accomplishing such analysis. A vast variety of nonlinear dimensionality reduction algorithms have been presented to preserve data's local structure including curvilinear components analysis [3], Sammon mapping [4], Isomap [5], Stochastic Neighbor Embedding [6], Laplacian Eigenmaps [7], Maximum Variance Unfolding [8], and Principal Component Analysis [9]. Despite their outstanding performance on simulated data sets, these approaches are generally ineffective at viewing actual, high-dimensional data. Most approaches, in particular, are incapable of keeping both the global and local structure of data in a single map [10]. Because only small pairwise distances are reliable in high-dimensional spaces, and in the low-dimensional embedding, most of these approaches simply aim to properly capture such short pairwise distances. Various studies have been interested in the topic of projecting multi-dimensional data into low dimensions because of its potential use to numerous types of data analytics. In this paper we choose to study two methods, t-distributed stochastic neighborhood embedding (t-SNE) and Least-Square Projection (LSP).

In 2008, Van Der Maaten et al. [11] presented visualized data by using t-SNE. They demonstrated "t-SNE," a novel approach for visualizing high dimensional data by assigning each data point a position based on a two or three dimensional plot. The method was a simplified version of Stochastic Neighbor Embedding, which creates substantially better visuals by minimizing the tendency for points to cluster together in the map's center. When it comes to constructing a single map that displays structure at several sizes, t-SNE outperforms previous approaches. This was especially significant for high - dimensional data that are also spread across numerous low dimension manifolds and yet were connected, such as photographs of items from various classes that can see from varied perspectives. They demonstrated in what way t-SNE can employ paths on neighborhood grids to enable the implicit pattern of all of the data to impact the way a portion of the data was shown for viewing the layout of really huge datasets. They also showed how t-SNE performs on a range of data sets and compared it to a number of other non-parametric visualization approaches, such as Locally Linear Embedding, Sammon mapping, and Isomap. In nearly all of the data sets, the visualizations created by t-SNE were much superior to those created by the other approaches.

In 2011 Van Der Maaten et al. [12] presented visualizing non-metric similarities in multiple maps. Multidimensional scaling approaches depict things like points on a low dimension metric map. They introduced an enhancement of the t-SNE multidimensional scaling approach, which was recently proposed. The goal of the modification was to solve the issues that classic multidimensional scaling approaches suffer when used to depict non-metric similarity. Several maps t-SNE, a novel approach, solves these issues by creating a number of maps that highlight complementing patterns in similar data. They used multiple mappings t-SNE to depict non-metric similarities in a huge data set of word association data and a data set of NIP'S co-authorship, proving its capacity to do so.

In the further advancement of the algorithm, Van Der Maaten et al. [13] investigated the accelerating of t-SNE—an embedding approach often used for the presentation of high dimension data in scatter plotting by applying two tree based algorithms. The research introduced Barnes Hut and dual-tree variations of algorithms that approach the gradients being used learning t-SNE embedding in $O(N \log N)$. Their studies found that the proposed algorithms significantly speed up t-SNE and enable the learning of embeddings for data sets containing millions of items. The Barnes-Hut t-SNE variant looked to outperform the dual-tree kind, which was quite perplexing.

Another advanced method of point-based visualization was put forward by Paulovich et al. [14] also in 2008. They developed a new multi-dimensional projecting approach that was based on the least squares approximation in their study. Relying on the parameters of a smaller number of control points having defined geometry, the estimations compute the parameters of a set of projected points. LSP sets the placement of their surrounding points based on a primary projection of the control points using a numeric solution that seeks to maintain a similar connection among the points indicated by a matrix. A minimal amount of distance computations was required to execute the projection, and no rearranging of the points was needed to attain final results with sufficient precision. The findings demonstrated the technique's capacity to group points in 2D based on their degree of similarities. They demonstrated this capacity by using it to map collections of textual texts from various sources, which was a strategic but challenging task. LSP outperformed other high-quality approaches in terms of speed and accuracy, especially where it was tested the most, particularly regarding mapping textual sets.

Principal Component Analysis (PCA) is used in this category of techniques to minimize the number of data dimensions. Rasmus et al [15] presented PCA in chemometrics, as well as a variety of other fields. It is one of the most important and powerful approaches, and it may be used for a variety of goals, such as demonstrating correlations between variables and samples (e.g., clustering), recognizing and quantifying patterns, and generating new ideas. Many latest studies [16], [17], [18], [19] have used these algorithms for different purposes. In this study, parameters of t-SNE and LSP are explored that are important for the understanding of mathematical notations impact. The impact of different parameters of t-SNE includes perplexity and iterations, and parameters of LSP include control point and the number of neighbors is shown on the images dataset. The impact of PCA parameters is also checked for the t-SNE implementation and represented in this study. Details of our methodology are given in the following section. R of our methodology as well as evaluation of the findings are given in the results and discussion section.

## Methodology

### Dataset

The COREL [20] dataset is used for visualizing the impact of mathematical notation changes using the algorithms. For a collection of images and drawings, Corel in the data archive is a vector space built of visual features. The data collection has ten labels and 1000 elements. These images include Africans, Beaches, Buildings, Buses, Dinosaurs, Elephants, Flowers, Food, Horses, and Mountains, and the size of each image is $256 \times 384$ or $384 \times 256$. LSP and t-SNE are applied to this dataset.

### t-Distributed Stochastic Neighbour Embedding (t-SNE)

t-SNE is a multidimensional projection technique that finds a lower dimensional representation of high dimensional dataset by minimising the kullback-Leibler (KL) divergence between the distribution that represents the high dimensional embedding space and the distribution that represents the corresponding lower dimensional embedding space. Given a high dimensional dataset $X = \{x_1, x_2, \dots, x_n\}$, we wish to find its lower-dimensional data representation $Y = \{y_1, y_2, \dots y_n\}$ of say of dimension $2\ or\ 3$ which can be displayed as a scatter plot for say visualization purposes. t-SNE begins by converting Euclidean distances between data points in high dimensional spaces into conditional probabilities that represent similarities. A Gaussian distribution around each data point in the high dimensional space is used to model such probabilities and is defined as follows in equation 1.

$$p_{j|i} = \frac{\exp(-\|x_i - x_j\|^2 / 2\sigma_i^2)}{\sum_{k \neq i} \exp(-\|x_i - x_k\|^2 / 2\sigma_i^2)} \quad (1)$$

where $p_{j|i}$ represents the conditional probability that $x_i$ would pick $x_j$ as its neighbor and $\sigma_i$ is the variance of the Gaussian distribution. t-SNE models the lower dimensional counterparts $y_i$ and $y_j$ of high dimensional points $x_i$ and $x_j$ as a student t-distribution around each data point as defined in equation 2.

$$q_{j|i} = \frac{(1 + \|y_i - y_j\|^2)^{-1}}{\sum_{k \neq l}(1 + \|y_k - y_l\|^2)^{-1}} \quad (2)$$

The choice of using the t-student distribution for the lower dimensional space is to mitigate the crowding effect associated with the Gaussian distribution.

A lower dimensional data representation is found by minimizing the mismatch between $p_{j|i}$ and $q_{j|i}$ with the KL divergence over all data points using gradient descent by equation 3.

$$C = KL(P||Q) = \sum_i \sum_j p_{ij} \log \frac{p_{ij}}{q_{ij}} \quad (3)$$

Where $C$ is the cost function, $P$ is the high-dimensional joint probability and $Q$ is the low-dimensional space. t-SNE's key parameter to be determined is the variance $\sigma_i$ of the Gaussian distribution. Due to the variability in the density of data points, there is no single optimal value of $\sigma_i$, and t-SNE uses a user specified perplexity value to search for the value of the variance with a joint probability representing a fixed perplexity value.

Perplexity can be seen as the effective number of neighbors for each data point by equation 4:

$$Perplexity(P_i) = 2^{H(P_i)} \quad (4)$$

where $H(P_i) = -\sum_j p_{j|i} \log_2 p_{j|i}$ is the Shannon entropy of the probability distribution.

The original paper suggest the optimal values for perplexity to be between 2 and 50, in this paper, we demonstrate and reveal the effect of different perplexity values on the quality of the generated projection.

**Principal Component Analysis (PCA)**

The classical PCA algorithm is linear and looks for a new basis to represent a high dimensional dataset by finding an affine subspace S of dimension $d \ll D$ to a set of centered points $\{x_1, \dots x_N\}$ in a high-dimensional space $R^D$. We begin by finding an orthogonal basis $v_1, \dots v_d$ for the data space such that the first principal component (PC) $v_1$ is the direction of the greatest variance of the data and the j-th PC $v_j$ is the direction orthogonal to the previous vectors $v_1, \dots, v_{j-1}$ of greatest variance.

The principal components are then used to find a low dimensional representation of the dataset by projecting the dataset onto the first d Principal components as follows:

From the 1st Principal component, we get a derived low dimensional representation as equation 5.

$$Y^{(1)} = v_1^T X \quad (5)$$

where $v_1 = \{v_{11}, \dots v_{1d}\}$ are chosen to maximize the Variance of $Y^{(1)}$. Generally, the i-th vector projection can be computed as follows by equation 6:

$$Y^{(i)} = v_i^T X \in R, \quad v_i \in R^D, i = 1, \dots, d \quad (6)$$

such that the variance of $y_i$ is maximized subject to equation 7.

$$v_i^T v_i = 1 \text{ and } Var(y_1) \geq Var(y_2) \geq \cdots \geq Var(y_d) > 0 \quad (7)$$

**Least Square Projection (LSP)**

Given a set of points $S = \{p_1, \ldots p_n\}$ in a vector space $R^m$, LSP technique finds a lower-dimensional representation of the points $R^d$ of S by persevering the neighborhood relationship as much as possible. LSP begins by projecting a subset of points in $S$ by any Multidimensional Scaling (MDS) method, these points are known as control points and constructs a system of linear equations which when solves gives the cartesian coordinates of the lower dimensional space points as a solution.

More formally, let $V_i = \{p_{i1}, \ldots, p_{ik_i}\}$ be considered as a set of $k_i$ points that are neighbours of a point $p_i$ and $\tilde{p}_i$ are the coordinates of the low dimensional space ($R^d$) of $p_i$ in $R^d$ defined by the following equation 8:

$$\tilde{p}_i - \sum_{p_j \in V_i} \alpha_{ij} \tilde{p}_j = 0, \quad (8)$$

where $0 \leq \alpha_{ij} \leq 1, \quad \sum \alpha_{ij} = 1$.

We then compute the coordinates of points $\tilde{p}_i$ from a set of linear systems by equation 9:

$$Lx_1 = 0, Lx_2 = 0, \ldots, Lx_d = 0 \quad (9)$$

where $x_1, \ldots, x_d$ are vector representations for the Cartesian co-ordinates of the points $(x_1, \ldots, x_n)$, $\alpha_{ij} = \frac{1}{k_i}$ when the point $p_i$ is in the centroid of points in $V_i$ and $L$ is the Laplacian matrix defined as follow by equation 10:

$$L = \begin{cases} 1, & i = j \\ -\alpha_{ij}, & p_j \in V_i \\ 0 & \text{otherwise} \end{cases} \quad (10)$$

LSP builds control points $S_c = \{p_{c_1}, \ldots, p_{c_{nc}}\}$ into a system of linear equations with the control points added as rows in the matrix and constructing a non-zero vector on the right hand side having coordintes of control points added. A system of linear equations can be now reformulated as follow by equation 11:

$$Ax = b \quad (11)$$

$A$ is defined as a rectangular $(n + nc) \times n$ matrix by the expression:

$$A = \begin{Bmatrix} L \\ C \end{Bmatrix}, \quad c_{ij} = \begin{cases} 1, & p_j \text{ is a control point} \\ 0, & \text{Otherwise} \end{cases}$$

and **b** is a vector:

$$b_i = \begin{cases} 0, & i \leq n, \\ x_{p_{c_i}}, & n < i \leq n + nc \end{cases}$$

Where $x_{p_{c_i}}$ defines one of the cartesian coordinates representing the control points $p_{c_i}$.

We obtain the solution of the linear equation by minimizing the expression $||Ax - b||^2$ using the method of least squares giving an exact solution given by $x = (A^T A)^{-1} A^T b$.

The key parameters under LSP are thus the a set of control points and the neighbors for each point. LSP finds the optimal group of control points by splitting the data set into $nc$ clusters utilising the k-medoids clustering method and then taking the medoid of each cluster as a control point. The defined control points are then projected into a lower-dimensional space $R^d$ through a Multidimensional scaling method and the initial layout is based on when interporating the remaining points to create a final layout.

The neighborhood points $V_i \in S$ for each point $p_i \in S$ are found using a clustering-based method, the clustering method finds the neighbourhood of a point by first searching the nearest neighbors of each cluster medoids and then only examining the cluster where $p_i$ belongs plus clusters that nearest to the examined cluster.

This work demystifies the parameters of LSP by showing how changing the number of control points and the number of neighbors selected by the users impacts the quality of the generated projections.

## Results and discussion

**Evaluation Metrics**

The projection quality of different hyperparameters of the projection was evaluated using the silhouette coefficient(Rousseeuw, 1987) and is computed by equation 12:

$$S = \frac{1}{n} \sum_{i=1}^{n} \frac{(b_i - a_i)}{max\{a_i, b_i\}} \qquad (12)$$

n represents the dimensionality of the features, considering a feature $i$, $a_i$ is the mean of the distances for features of the same class, $b_i$ accounts for the least mean distance between all features with other features of a different class. The coefficient ranges from -1 to 1 and the closer to 1, the better the projection quality.

We also used the neighborhood hit ( graph to assess the quality of projection for different hyperparameters. It shows the average proportion of points in a neighborhood that belond to the same label. The graph is constructed increasing the neighborhood and seeing how that quantity progresses.

**Mathematical impacts on projections**

We have used a different ranges of parameters and visualized the impact of their impact in the result of the projection.. For the LSP, we have changed the number of control points [25, 50, 75] and the number of neighbors [10, 20] and calculated the silhouette score for them. The score for all combinations is shown in Table 1.

Table 1: LSP parameters and their effect on model performance

| Number of control points (CP) | Number of neighbors (NN) | Silhouette score |
|---|---|---|
| 25 | 10 | 0.36 |
| 25 | 20 | 0.37 |
| 50 | 10 | 0.29 |
| 50 | 20 | 0.34 |
| 75 | 10 | 0.23 |
| 75 | 20 | 0.26 |

Through these experiments, we can see that as the number of control points increases, the performance of the model is decreasing. We have shown the neighborhood hit in Figure 1.

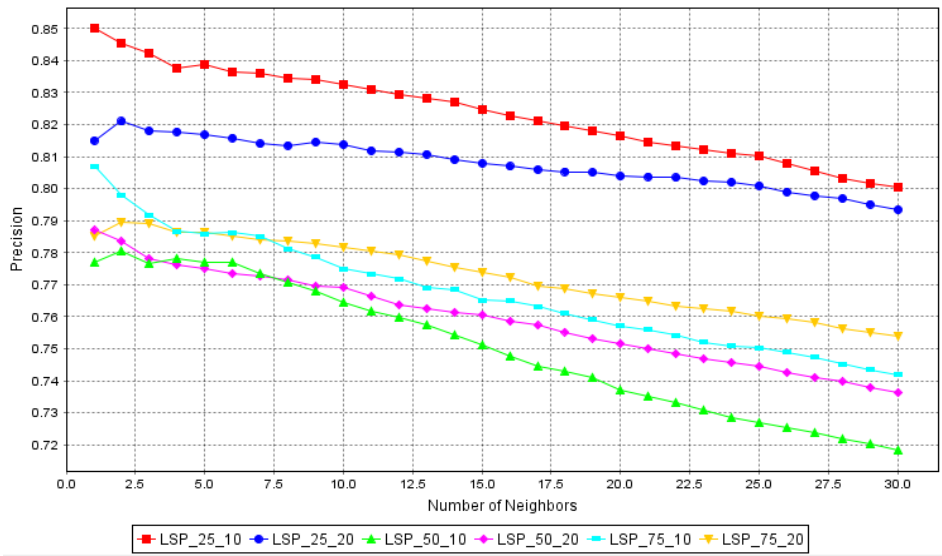

Figure 1: LSP neighborhood hit comparison chart

In the caption of the image, the first number is the number of control points and the second is the number of neighbors. Precision is decreasing as the number of control points is increasing. For the best parameters of LSP (CP:25 and NN:20) the projection is shown in Figure 2 while the parameters with lowest values (CP:75 and NN: 10) are shown in Figure 3.

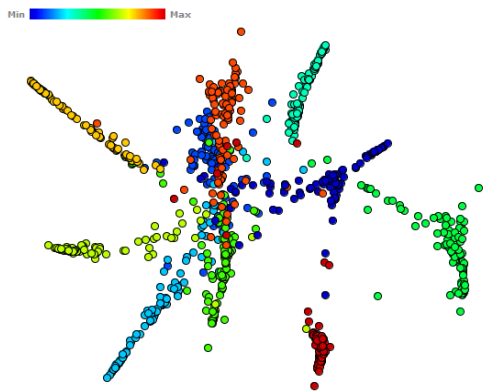     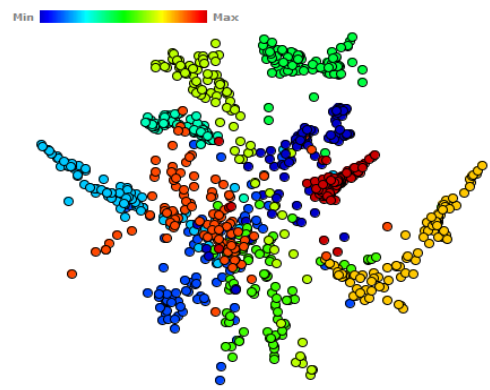

Figure 2: LSP least score projection      Figure 3: LSP best score projection

In these images, each point shows an actual data point, in figure 2 data are not overlapping and separated batter than in figure 3.

For the t-SNE, we have used the perplexity in the range [20, 30, 40] and the number of iterations [1000, 1500] and calculated the silhouette score for them. The score for all combinations is shown in table 2.

Table 2: t-SNE parameters and their effect on model performance

| Perplexity | Number of Iterations | Silhouette Score |
|---|---|---|
| 20 | 1000 | 0.46 |
| 30 | 1000 | 0.42 |
| 40 | 1000 | 0.45 |
| 20 | 1500 | 0.47 |
| 30 | 1500 | 0.47 |
| 40 | 1500 | 0.46 |

These experiments show that number of iterations is not much effective while perplexity has a significant effect on the performance of the model. The neighborhood hit comparisons for these parameters are shown in Figure 4.

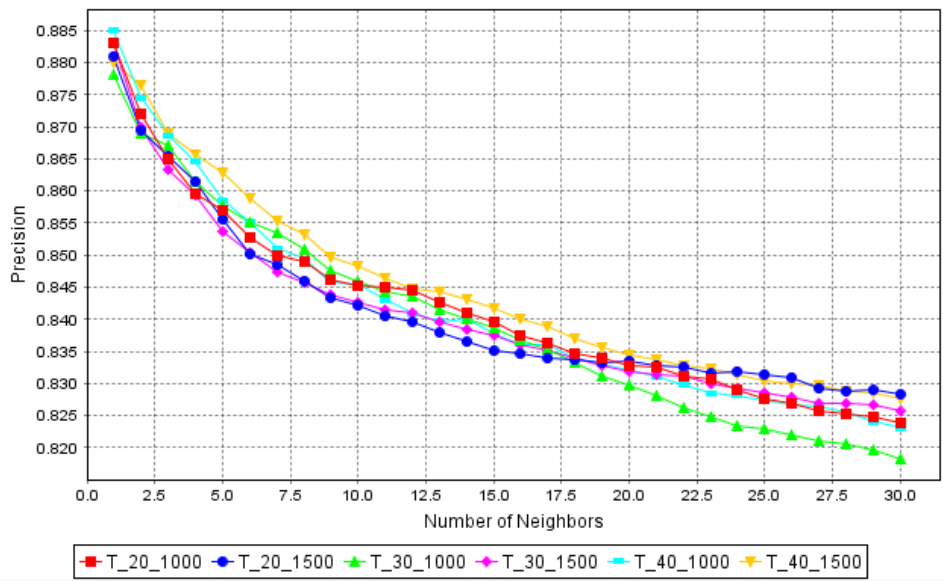

Figure 4: t-SNE neighborhood hit comparison chart

There is not much gap between the progress of models with the different number of parameters. The projection of best parameter values (perplexity:30 and iterations:1500) is shown in figure 5 while with least score (perplexity:30 and iterations:1000) is shown in figure 6.

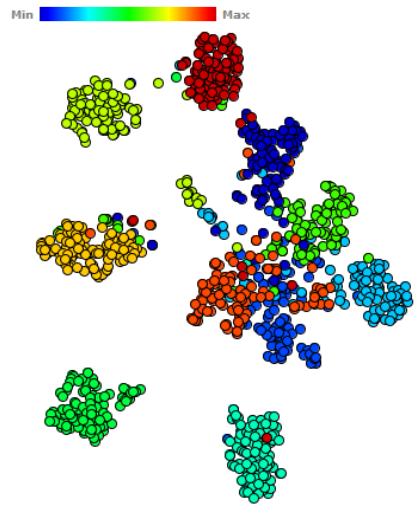

Figure 5: t-SNE least score projection

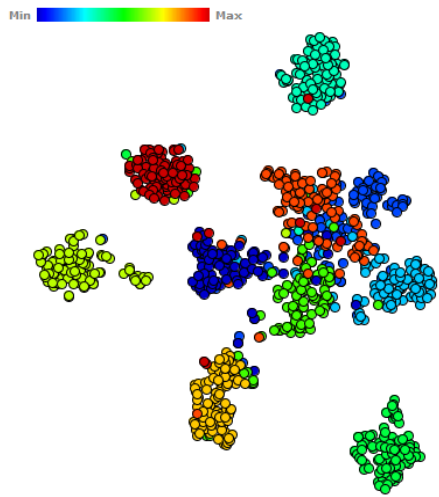

Figure 6: t-SNE best score projection

There is a significant difference between the LSP and t-SNE score and neighborhood score, we have shown that in image 7.

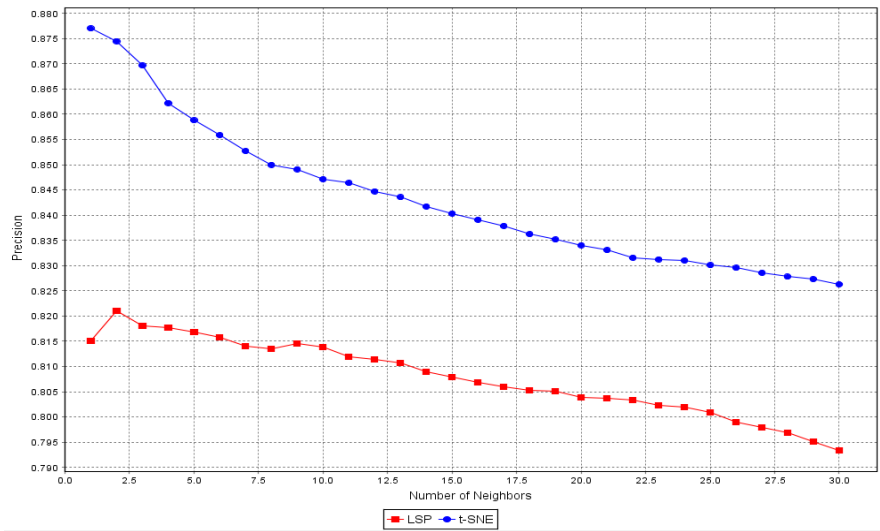

Figure 7: neighborhood hit comparsion between LSP and t-SNE

In the tables and image 7, there is a significant difference between the performance of LSP and t-SNE, and t-SNE is performing better than LSP as far as segregation of labels is concerned. That is also reflected by the visualizations. As t-SNE is giving better results and to know the impact of PCA behind that, the different number of features were checked at the best parameter set of t-SNE. The range and results of these parameters are given in Table 3.

| PCA Features | Perplexity | Number of Iterations | Silhouette Score |
|---|---|---|---|
| 20 | 30 | 1500 | 0.46 |
| 30 | 30 | 1500 | 0.47 |
| 40 | 30 | 1500 | 0.49 |

t-SNE is giving the best score with feature space size 40, selected by PCA. The projection of the best parameters is shown in figure 8 and the neighborhood hit score is shown in Figure 9.

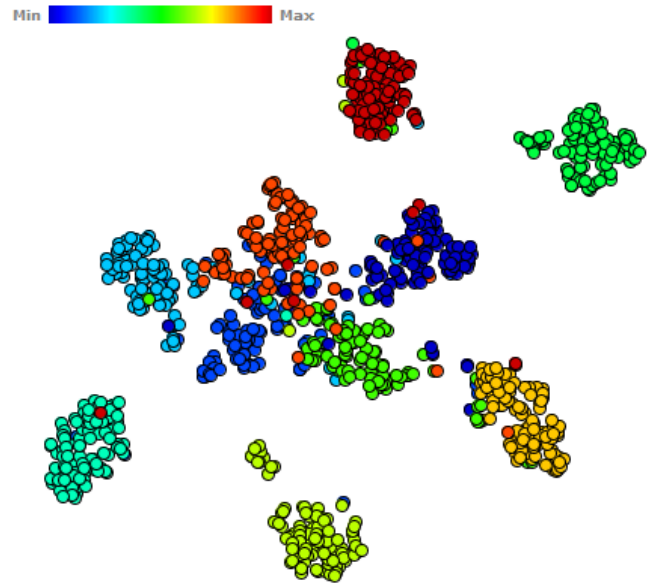

Figure 8: PCA tuned t-SNE projection

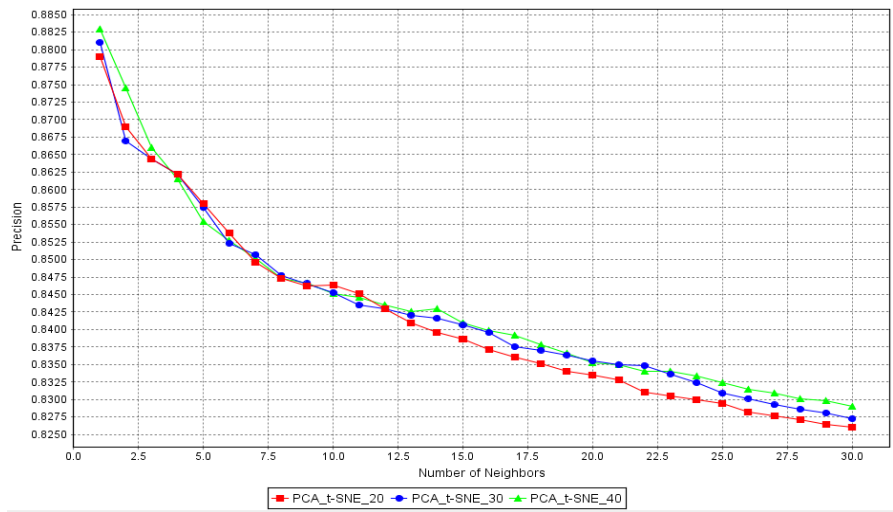

Figure 9: PCA tuned t-SNE neighborhood hit score

These images show the impact of feature selection using PCA. The more features are giving the batter results not just in silhouette but also in the neighborhood hit.

## Conclusion

Data visualization plays an important role in understanding the complex data structure, mathematics, and effect of different mathematical variations in the real world. In this study, we have discussed and presented the impact of mathematical parameters on the visualization. The Corel images dataset and t-SNE and LSP are used to demonstrate how the changing in mathematical parameters can affect the visualization. Data is clustered visualized using the point-based data clustering algorithm and separated the data into similar groups. These algorithms are based on core mathematical notations and evaluated by different kinds of distance measures. We have used a different range of parameters in this study and looked at the impact of these parameters by using evaluation measures and projections. These projections help us to see how well data is clustered. By doing a different range of experiments like changing the LSP parameters that include the number of control points and the number of neighbors, and t-SNE parameters that include perplexity and number of iterations, and PCA behind the t-SNE, we were able to achieve the 0.49 silhouette score which is good in the context of clustering the data. In order to be completely fair in the performance comparison between LSP and t-SNE, though, we are preparing an experiment to run LSP after PCA is performed on the original data set.